\documentclass[%
 aip,
 amsmath,amssymb,
 reprint,%
citeautoscript              % Automatically places superscript citations /after/ punctuation
]{revtex4-1}
\usepackage{color}
\usepackage[usenames,dvipsnames,svgnames,table]{xcolor}
\usepackage[colorlinks=true,linkcolor=blue,urlcolor=blue,citecolor=blue]{hyperref}
\usepackage{mathtools}
\usepackage{float}
\usepackage{graphicx}
\usepackage{dcolumn}
\usepackage{array}
\usepackage{lipsum}
\usepackage{bm}
\usepackage{subfigure}
\usepackage{amssymb}
\usepackage{multirow}
\usepackage{tabularx}
\usepackage{amsmath}
\usepackage{braket}
\usepackage{csquotes}
\graphicspath{{plots/}}
 \usepackage{lipsum}
\usepackage{mathrsfs}
\usepackage{MnSymbol}
\usepackage[capitalize]{cleveref}
\crefname{figure}{fig.}{figs.}
\Crefname{figure}{Fig.}{Figs.}
\crefname{table}{tab.}{tabs.}
\Crefname{table}{Tab.}{Tabs.}
\crefname{equation}{eq.}{eqs.}
\Crefname{equation}{Eq.}{Eqs.}
\crefname{section}{sec.}{secs.}
\Crefname{section}{Sec.}{Secs.}
\newcommand{\beq}{\begin{equation}}
\newcommand{\eeq}{\end{equation}}
\newcommand{\bea}{\begin{eqnarray}}
\newcommand{\eea}{\end{eqnarray}}

\newcommand{\unitspace}{\hspace{2pt}} % Use in math-mode as the space between units, e.g. g cm^-3

\newcommand{\lap}[1]{\mathcal{L}[#1]}

\newcommand{\dom}{\,\mathrm{d}\omega}

\newcommand{\mc}[1]{\mathcal{#1}}
\newcommand{\mbf}[1]{\mathbf{#1}}

\usepackage{booktabs}
\usepackage{tikz}
\usepackage{pgfplots}
\pgfplotsset{compat=1.18} 
\usepackage[linesnumbered,ruled,vlined]{algorithm2e}
\usepackage[normalem]{ulem}
\begin{document}
%\underline{Draft version:}
\title{Correlation function metrology for warm dense matter: Recent developments and practical guidelines}

\author{M.~P.~B\"ohme}
\email{boehme4@llnl.gov}
\affiliation{Quantum Simulations Group, Physics and Life Science Directorate, Lawrence Livermore National Laboratory (LLNL), California 94550 Livermore, USA}

\author{W.~M.~Martin}
\affiliation{Physics Department, Stanford University, Stanford, California 94305, USA}
\affiliation{SLAC National Accelerator Laboratory, Menlo Park California 94309, USA}

\author{H.~M.~Bellenbaum}
\affiliation{Center for Advanced Systems Understanding (CASUS), D-02826 G\"orlitz, Germany}
\affiliation{Helmholtz-Zentrum Dresden-Rossendorf (HZDR), D-01328 Dresden, Germany}
\affiliation{Institut f\"ur Physik, Universit\"at Rostock, D-18057 Rostock, Germany}

\author{M.~Berrens}
\affiliation{Quantum Simulations Group, Physics and Life Science Directorate, Lawrence Livermore National Laboratory (LLNL), California 94550 Livermore, USA}

\author{J.~Vorberger}
\affiliation{Helmholtz-Zentrum Dresden-Rossendorf (HZDR), D-01328 Dresden, Germany}

\author{S.~Schwalbe}
\affiliation{Center for Advanced Systems Understanding (CASUS), D-02826 G\"orlitz, Germany}
\affiliation{Helmholtz-Zentrum Dresden-Rossendorf (HZDR), D-01328 Dresden, Germany}

\author{Zh.~A.~Moldabekov}
\affiliation{Center for Advanced Systems Understanding (CASUS), D-02826 G\"orlitz, Germany}
\affiliation{Helmholtz-Zentrum Dresden-Rossendorf (HZDR), D-01328 Dresden, Germany}

\author{Th.~Gawne}
\affiliation{Center for Advanced Systems Understanding (CASUS), D-02826 G\"orlitz, Germany}
\affiliation{Helmholtz-Zentrum Dresden-Rossendorf (HZDR), D-01328 Dresden, Germany}

\author{S.~Hamel}
\affiliation{Quantum Simulations Group, Physics and Life Science Directorate, Lawrence Livermore National Laboratory (LLNL), California 94550 Livermore, USA}

\author{B.~Aguilar-Solis}
\affiliation{University of California, Merced, 5200 North Lake Road, Merced, CA 95343, USA}

\author{A.~Sharma}
\affiliation{Department of Civil Engineering, Indian Institute of Technology Roorkee, Roorkee, India}

\author{F.~R.~Graziani}
\affiliation{NIF and Photon Science, Lawrence Livermore National Laboratory (LLNL), 94550 Livermore California, USA}

\author{T.~D\"oppner}
\affiliation{NIF and Photon Science, Lawrence Livermore National Laboratory (LLNL), 94550 Livermore California, USA}

\author{S.~H.~Glenzer}
\affiliation{SLAC National Accelerator Laboratory, Menlo Park California 94309, USA}

\author{T.~Dornheim}
\affiliation{Center for Advanced Systems Understanding (CASUS), D-02826 G\"orlitz, Germany}
\affiliation{Helmholtz-Zentrum Dresden-Rossendorf (HZDR), D-01328 Dresden, Germany}

\author{D.~T.~Bishel}
\email{bishel1@llnl.gov}
\affiliation{NIF and Photon Science, Lawrence Livermore National Laboratory (LLNL), 94550 Livermore California, USA}

\begin{abstract}
X-ray Thomson scattering (XRTS) has emerged as a valuable diagnostic for matter under extreme conditions, as it captures the intricate many-body physics of the probed sample. Recent advances, such as the model-free temperature diagnostic of Dornheim \textit{et al.}~[Nat.~Commun.~\textbf{13}, 7911 (2022)], have demonstrated how much information can be extracted directly within the imaginary-time formalism. However, since the imaginary-time formalism is a concept often difficult to grasp, we provide here a systematic overview of its theoretical foundations and explicitly demonstrate its practical applications to temperature inference, including relevant subtleties. Furthermore, we present recent developments that enable the determination of the absolute normalization, Rayleigh weight, and density from XRTS measurements without reliance on uncontrolled model assumptions. Finally, we outline a unified workflow that guides the extraction of these key observables, offering a practical framework for applying the method to interpret experimental measurements.
\end{abstract}

\maketitle

\section{Introduction}

Diagnostics of dense plasmas %provide a unique window
%into
are of paramount importance to understand the microscopic states of matter at extreme conditions where Coulomb coupling, partial ionization, and quantum degeneracy all play decisive roles \cite{vorberger2025roadmapwarmdensematter,Bonitz_POP_2024,drake2018high}. In this non-ideal warm dense matter (WDM) regime, many standard approximations break down, so robust constraints from experiment are essential. X-ray Thomson scattering (XRTS) has become a premier probe for this purpose, enabling access to collective and single-particle excitations and, e.g., through detailed-balance, to thermodynamic information \cite{siegfried_review,kraus_xrts,DOPPNER2009182,Gregori_PRE_2003}. 

The central observable in XRTS is the electronic dynamic structure factor (DSF), which represents the correlations of density fluctuations within the system. Current experimental platforms have enabled the measurement of XRTS spectra %on both inertial confinement fusion (ICF)
using both backlighter X-ray sources~\cite{MacDonald_POP_2022,Tilo_Nature_2023,Glenzer_PRL_2007,shi2025firstprinciplesanalysiswarmdense} and X-ray free electron lasers (XFELs) within the last two decades \cite{Fletcher2013,Gawne_PRB_2024,Martin_POP_2025,Bellenbaum_APL_2025,bespalov2025experimentalvalidationelectroncorrelation}. Experimental spectra are typically interpreted via forward modeling, in which a model combining theoretical predictions and experimental effects is fitted to the measured data~\cite{KasimXRTS}. One of the most commonly used theories to model the DSF is the Chihara decomposition \cite{Gregori_PRE_2003,Chihara_1987,bohme2023evidencefreeboundtransitionswarm}. The primary ansatz of the Chihara decomposition is the separation of the electronic density operator into bound and free contributions to compute a DSF for a given charge state, density and temperature \cite{Wunsch2011,Chapman2014,Chapman_2015,BellenbaumIonization}. This ansatz requires precise knowledge of the microscopic electronic states in the plasma \cite{MatternSeidler2013}. Chihara models are often based on approximations within ground state theories or within the ideal plasma regime that have not been sufficiently tested for application to WDM. The approximations can therefore be inaccurate at thermal energies of several eVs and high compression rates, where excited states and non-ideal effects can be expected.

One prominent example is the effect of ionization potential depression (IPD), which refers to the lowering of the effective binding energy of bound states in a plasma environment. Several experimental studies have shown that effective ionization energies under WDM conditions are not predicted accurately \cite{ciricosta_2016,Hoarty2013,Crowley2014,Fletcher2014,Iglesias2014,kraus_xrts} by commonly used semi-empirical models such as the Ecker-K\"oll \cite{Ecker1963} or Stewart-Pyatt models \cite{Stewart1966}. While the models include most of the relevant physics, their main limitations lie in determining the relative strength of screening and in correctly describing the crossover between different screening descriptions. To expand our understanding of the behavior of bound electrons under extreme conditions, first principles studies have been conducted using advanced techniques such as average atom models~\cite{Massacrier2021} (AA) and density functional theory calculations \cite{Hu2016,PhysRevResearch.2.023260,gawne2023investigating}. While recent path-integral Monte Carlo (PIMC) studies for the case of hydrogen~\cite{Bonitz_POP_2024,Bonitz_CPP_2025,BellenbaumIonization} have extracted ionization energies from quasi-exact simulations, the accessible system sizes  remain limited to tens of ions due to PIMC's exponential computational costs \cite{dornheim_sign_problem}. This example demonstrates the difficulties in the modeling of experiments in the WDM regime.

Recently, a complementary route formulated in the imaginary-time domain has proven especially useful~\cite{Dornheim_T_2022,Dornheim2022Physical,Dornheim_PTR_2023}. By recasting the problem in terms of the imaginary-time correlation function (ITCF), which is the Laplace transform of the dynamic structure factor, one can exploit the numerical stability of Laplace space and directly leverage equilibrium symmetry \cite{Dornheim_T_2022,Thermometry2023,Dornheim2022Physical}. This has led to a model-free temperature diagnostic and reliable uncertainty quantification even for noisy data \cite{Thermometry2023,shi2025firstprinciplesanalysiswarmdense,Schoerner_PRE_2023,Dornheim2025-sv,Bellenbaum_APL_2025}. Beyond temperature, the imaginary-time framework provides a natural path to the \textit{a priori} unknown absolute normalization of the measured XRTS spectrum via the universal $f$-sum rule~\cite{dornheim2023xray}. This, in turn, gives one access to the electronic static structure factor $S_{ee}(k)$, opening the door to additional observables such as the Rayleigh weight and static density response, all of which can be directly related to the measured spectra \cite{Thermometry2023,schwalbe2025staticlineardensityresponse,dornheim2024modelfreerayleighweightxray}. 
The ITCF framework has already been used to refine physical modeling (e.g., identifying free–bound signatures relevant for equation-of-state measurements \cite{bohme2023evidencefreeboundtransitionswarm}) and to detect out-of-equilibrium states \cite{VORBERGER2024129362}.
At the same time, progress on forward modeling and instrument characterization continues to improve the fidelity of the combined source-and-instrument function (SIF) and thus the robustness of any remaining deconvolution step \cite{Gawne2024,Gawne_PRB_2024}. Together, these developments delineate a correlation-function–centric metrology for WDM that is both mathematically rigorous and broadly applicable across experimental platforms.

Although this method is both powerful and straightforward to implement, its origin in first-principles quantum Monte Carlo studies \cite{cep,Filinov2016,Boninsegni1996,Dornheim2022Physical,Dornheim_JCP_ITCF_2021} may make it less accessible to a broad audience. To encourage wider adoption, we take a pedagogical approach: we review the method’s origins and provide a worked example that extracts temperature from an experimental spectrum using the imaginary-time formalism. We also introduce absolute normalization via the f-sum rule, enabling inference of the electronic static structure factor $S_{ee}(\mathbf{k})$ and, in turn, the Rayleigh weight $W_R(\mathbf{k})$. In addition, we present a diagnostic for non-equilibrium conditions in XRTS measurements. Overall, this work offers a methodological synopsis of imaginary-time techniques in the WDM context.

\begin{figure}
    \centering
    \includegraphics[width=0.485\textwidth]{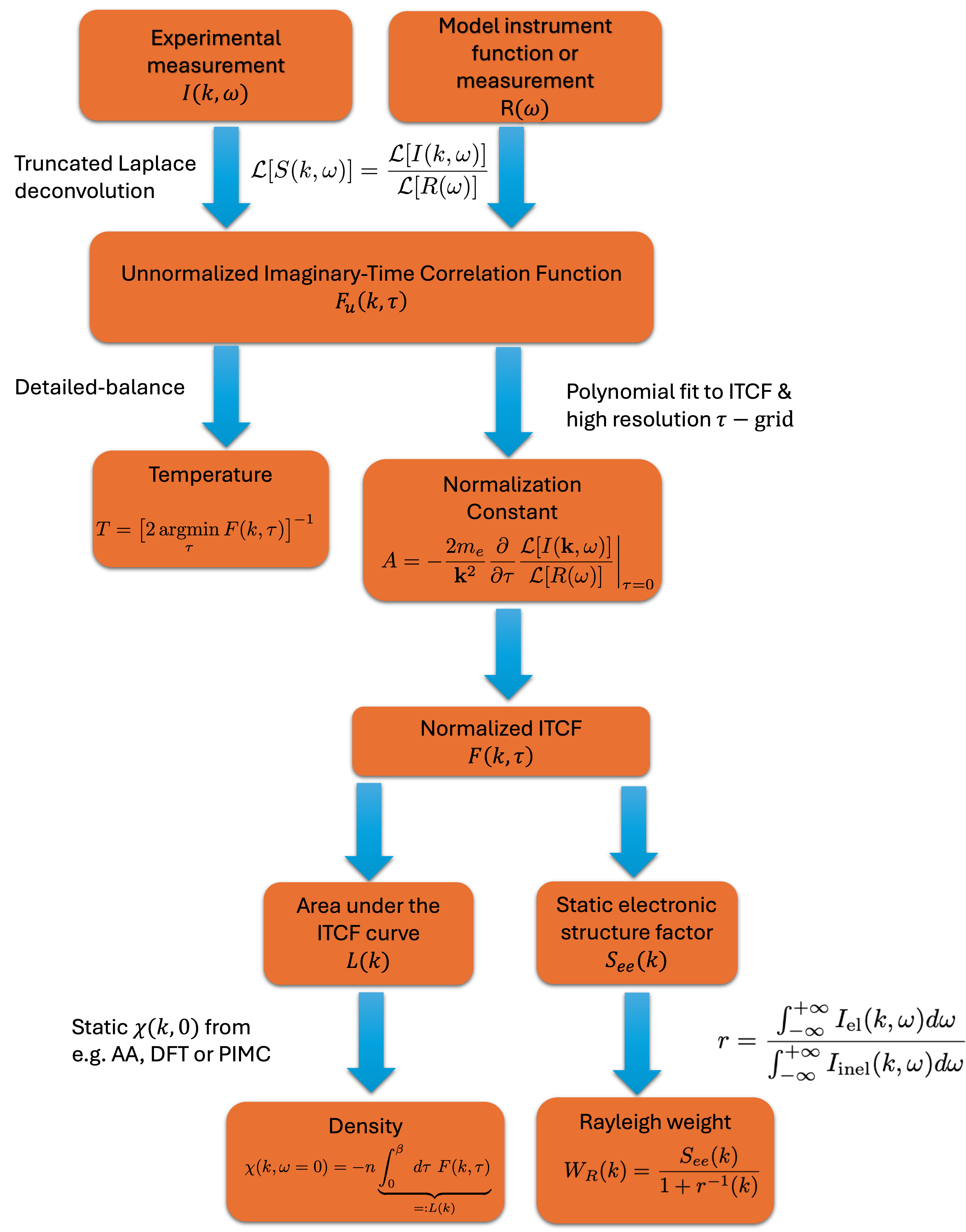}
    \caption{Workflow for correlation-function metrology from XRTS. Starting from the measured intensity $I(\mbf{k},\omega)$ and a measured or modeled source\textendash and\textendash instrument function (SIF) $R(\omega)$, a truncated two-sided Laplace deconvolution yields the (unnormalized) imaginary-time correlation function (ITCF) $F_u(\mbf{k},\tau)$. Detailed-balance symmetry fixes the temperature via the location of the minimum at $\tau=\beta/2$. The $\tau\!\to\!0$ slope, through the $f$-sum rule, provides the absolute normalization. The normalized ITCF supplies $S_{ee}(\mbf{k})=F(\mbf{k},0)$ and, via \Cref{eq:FDT}, access to the static density response. Combining $S_{ee}(\mbf{k})$ with the elastic/inelastic ratio gives the Rayleigh weight $W_R(\mbf{k})$, while comparison of $\chi(\mbf{k},0)$ to theoretical benchmarks enables density inference. Convergence with respect to the deconvolution window and the $\tau$-grid, together with accurate SIF characterization, are the only analysis prerequisites.}
    \label{fig:Flowchart}
\end{figure}

We first provide an overview of the methods used in this work and their connection to the experimental measurements. Figure (\ref{fig:Flowchart}) summarizes the workflow developed in Refs.~\cite{Dornheim_T_2022,Thermometry2023,schwalbe2025staticlineardensityresponse,dornheim2023xray,Bellenbaum_APL_2025,dornheim2024modelfreerayleighweightxray,VORBERGER2024129362,Dornheim2025-sv}~. We start from the measured intensity $I(\mathbf{k},\omega)$ together with a model or measurement of the SIF $R(\omega)$; the theoretical foundations of the measurement process are reviewed in \Cref{ssec:xrts}. Next, we obtain the deconvolved, unnormalized ITCF from $I(\mathbf{k},\omega)$ and $R(\omega)$. The methodology is introduced in \Cref{ssec:ITCF} and illustrated on an example measurement in \Cref{ssec:temp_xfel} based on Ref.~\cite{Martin_POP_2025}. Even without normalization, the ITCF $F_u(\mathbf{k},\tau)$ enables temperature estimation and diagnosis of non-equilibrium when scattering is measured at multiple angles, as discussed in \Cref{ssec:noneq}. To exploit the full diagnostic power of the ITCF, proper normalization is essential. We reiterate the theoretical basis in \Cref{ssec:fsum} from Ref.~\cite{dornheim2023xray} and show how it enables computation of the static structure factor $S_{ee}(\mathbf{k})$. This, in turn, allows extraction of the Rayleigh weight $W_R(\mathbf{k})$, presented in \Cref{ssec:Wr}. $W_R$ provides insight into electron localization and enables inference of the sample density from first principles \cite{Dornheim2025-sv}. A complementary approach based on the imaginary-time fluctuation-dissipation theorem, described in \Cref{ssec:area}, requires only the area under the normalized ITCF $L(\mbf{k})$ \cite{schwalbe2025staticlineardensityresponse}. Overall, the workflow demonstrates how to obtain the key observables $T$, $S_{ee}(\mathbf{k})$, and $W_R(\mathbf{k})$ from first principles without uncontrolled model assumptions, provided the SIF $R(\omega)$ is accurately known. To infer the density $\rho$, the workflow requires a comparison against theoretical calculations and outlines two independent ways of inferring the density\cite{schwalbe2025staticlineardensityresponse,Dornheim2025-sv}. This enables a consistency check of the theories and increases the robustness of the approach.

In summary, \Cref{sec:theory} gives the reader a brief overview of the origin of imaginary-time methods and their relation to measurements in real frequency space. The main part of this work is presented in \Cref{sec:results}, where we provide a detailed example of how to compute the ITCF, along with a discussion of the associated subtleties and potential pitfalls. We conclude in \Cref{sec:summary} with an outlook on leveraging these developments for multidiagnostic and multi-angle extensions \cite{Bellenbaum_APL_2025}. We aim to promote a better understanding of the ITCF method as a reliable means of characterizing the full plasma state.

\section{Theory\label{sec:theory}}

In this section, we give an overview of the DSF, which governs the physics of XRTS. We review the physical measurement process to show the difference between the DSF and the recorded spectrum. In the last part of this section, we focus on the intricate connection between the DSF and the imaginary-time formalism. All the following derivations are done in Hartree atomic units ($\hbar = m_e = e = 4\pi \varepsilon_0$ = 1).
\subsection{X-ray Thomson scattering}
\label{ssec:xrts}

%The usage of imaginary time for experimental inference has been a novel development pioneered by Dornheim et al.~\cite{Dornheim_T_2022}. %Using the imaginary-time formalism, we can obtain the exact temperature of the probed sample from an XRTS measurement without any model assumption if the probing region is homogeneous and in a state of thermal equilibrium. We will review the detection of nonequilibrium in \Cref{sec:results}. 

XRTS experimental setups consist of an X-ray source, commonly a laser-driven foil or an XFEL beam, that probes a sample with the wavevector $\mathbf{k}_i$ and a detector that records the final wave-vector  $\mathbf{k}_s$. The scattering wave vector is given by $\mathbf{k} = \mathbf{k}_s - \mathbf{k}_i$. 
The process of a photon getting scattered with a wave vector $\mathbf{k}$ and a frequency $\omega$ is described by the differential cross-section \cite{Chihara_1987}
\begin{equation}
    \frac{\partial^2 \sigma}{\partial \Omega \partial \omega}(\mathbf{k},\omega) = \sigma_T \frac{k_s}{k_i} S_{ee}(\mathbf{k}, \omega), 
    \end{equation}
with $\sigma_T= 6.65 \times 10^{-29}$ m$^2$ the total Thomson cross-section and $S_{ee}(\mathbf{k},\omega)$ the electronic dynamical structure factor of the system. In this paper, we focus on non-relativistic, quasi-elastic scattering measured at a constant scattering angle. These assumptions are appropriate for most XRTS experiments on WDM but break down for optical Thomson scattering and X-ray energies of 10 keV and more. The DSF is a quantity that represents the equilibrium density fluctuations of the system. To be precise, it is the Fourier transform of the intermediate scattering function for a system consisting of two particle species $a$ and $b$:
\begin{equation}\label{eq:DSF_FT}
    S_{ab}(\mathbf{k},\omega) = \frac{1}{2\pi} \int_{-\infty}^{+\infty} \ \textnormal{d}t \ F_{a b}(\mathbf{k}, t) e^{i \omega t},
\end{equation}
with the intermediate scattering function defined as
\begin{equation}\label{eq:intermediate}
F_{a b}(\mathbf{k}, t)=\frac{1}{\sqrt{N_a N_b}}\left\langle\hat{\rho}_a(\mathbf{k}, t) \hat{\rho}_b^*(\mathbf{k}, 0)\right\rangle.
\end{equation}
One should note that the average in \Cref{eq:intermediate} is to be understood as the thermal average within the canonical ensemble as
\begin{equation}\label{eq:Th_avg}
    \langle A \rangle = \text{Tr}\left[\frac{e^{-\beta \hat{H}}}{Z} \hat{A} \right],
\end{equation}
where $Z$ is the canonical partition function, $\hat{H}$ represents the many-body Hamiltonian of the system, $\beta = (k_BT)^{-1}$ the inverse thermal energy and $\text{Tr}[\dots]$ is the trace operator. We review the canonical partition function more in-depth in the next subsection. 
$\rho_{a}(\mathbf{k},t)$ are the particle density operator in the Heisenberg picture for particles $a$
\begin{equation}\label{eq:density_rt}
    \hat{\rho}_{a}(\mathbf{k},t) = e^{i\hat{H}t} \left[ \sum_{i=1}^{N _a} e^{i\mathbf{k} \mathbf{r}_i} \right] e^{-i\hat{H}t}.
\end{equation}
In the case of  the electronic dynamic structure factor, the density operators are commonly split up into contributions from bound and free electrons
\begin{equation}
    \rho(\mathbf{k})= \rho_\text{bound}(\mathbf{k}) + \rho_{\text{free}}(\mathbf{k}),
\end{equation}
which culminates in the Chihara decomposition \cite{Chihara_1987} when applying this ansatz to \Cref{eq:intermediate}. One should note that the splitting of the electrons into bound and free populations is often difficult and not unique in the case of dense plasmas as orbital overlaps can make this decomposition ambiguous. This is caused by the resulting delocalization of bound electronic states, which makes these almost indistinguishable from continuum states. 

One important insight is gained through an alternative definition of the DSF, the so-called spectral decomposition \cite{quantum_theory,Sturm+1993+233+242}
\begin{equation}
\label{eq:DSF_spectral_rep}
    S(\mathbf{k},\omega) = \frac{1}{Z}\sum_{n,m} e^{-\beta E_m} |\bra{n}\hat{\rho}(\mathbf{k})\ket{m}|^2 \delta(\omega - \omega_{mn} ) \ ,
\end{equation}
where $n,m$ refers to a sum over all the possible many-body eigenstates of the full Hamiltonian, $E_m$ the eigenenergy of the m-th eigenstate of $\hat{H}$ and $\omega_{nm}$ the energy difference between the states. By exchanging $n \rightleftharpoons m$, inserting  $1 = e^{\beta \omega} \ e^{-\beta \omega}$, and using the evenness in \Cref{eq:DSF_spectral_rep} of the delta-function, we arrive at the detailed balance relation \cite{quantum_theory}
\begin{equation}
\label{eq:detailed_balance_dsf}
    S(\mathbf{k},-\omega) = e^{-\beta \omega} S(\mathbf{k},\omega) \ .
\end{equation}
The above relation reveals a central symmetry in the theory of equilibrium density fluctuations that, in principle, we can utilize to obtain a temperature of any XRTS measurement~\cite{DOPPNER2009182}. 

In practice, this is generally not directly applicable, as the measurement process must be performed using an X-ray beam of finite bandwidth and broadening from the instrument. As a result, the detector does not record the DSF itself, but rather its convolution with the SIF $R(\omega)$ \cite{Gawne_JAP_2024,Gawne_PRB_2024,gawne2025heartnewxraytracing}
\begin{equation}
    I(\mathbf{k},\omega) = S(\mathbf{k},\omega) \circledast R(\omega) = \int_{-\infty}^{\infty} d\omega' S(\mathbf{k},\omega-\omega') R(\omega').
\end{equation}
The SIF accounts for the spectral content of the source and broadening effects of the detection instrument, such as the mosaicity of a crystal analyzer~\cite{Gawne_JAP_2024}. To truly infer temperature from \Cref{eq:detailed_balance_dsf} one must first deconvolve the DSF from the measured signal. %This is extremely difficult in practice, as the presence of experimental noise makes common deconvolution techniques such as the Fourier deconvolution numerically unstable. 
\subsection{Origin of imaginary-time methods}

Since the ITCF method \cite{Dornheim_T_2022} is inspired by the imaginary-time path-integral Monte Carlo (PIMC) method~\cite{cep,Berne_JCP_1982,Takahashi_Imada_PIMC_1984}, we wish to give the reader a brief overview on the basic concepts that define imaginary time. 

The real-time propagator of a system with a time-independent Hamiltonian $\hat{H}$ is defined as
\begin{align}
    K(x_a,t_a;x_b,t_b) = \braket{x_a| e^{-i\hat{H}(t_a-t_b)}|x_b},
\end{align}
for a system to transition from coordinate $x_a$ at time $t_a$ to $x_b$ at $t_b$. 
If we now take the propagator that computes the conditional probability of a system transition $\ket{x_a} \rightarrow \ket{x_b}$ in between times 0 and $t$, we have 

\begin{equation}
    K(x_a,t; x_b, 0) = \bra{x_a} e^{-i\hat{H}t} \ket{x_b}.
\end{equation}
We can now perform a Wick rotation $t \rightarrow -i\tau$  with $\tau \in [0,\beta]$ of this expression and obtain the so-called \textit{imaginary-time} propagator

\begin{equation}
    K(x_a,\tau;x_b,0) = \bra{x_a} e^{-\tau \hat{H}}\ket{x_b}.
\end{equation}
The canonical partition function 
\begin{equation} \label{eq:Part_function}
    Z  = \int dx \ \bra{x} e^{-\tau\hat{H}}\ket{x}\Big|_{\tau=\beta} = \text{Tr}\left[e^{-\beta\hat{H}} \right], 
\end{equation}
is obtained by summing over all possible positions in the configuration space for distinguishable particles at temperature $\beta = (k_B T)^{-1}$ for a system transitioning back into its original state.
Thus, for equilibrium systems, the real-time dynamics of closed paths connects to the statistical averages of the system. This fact is further discussed in Ref.~\cite{Stefanucci_van_Leeuwen_2025} . 

\begin{figure}
    \centering
    \includegraphics[width=0.5\textwidth]{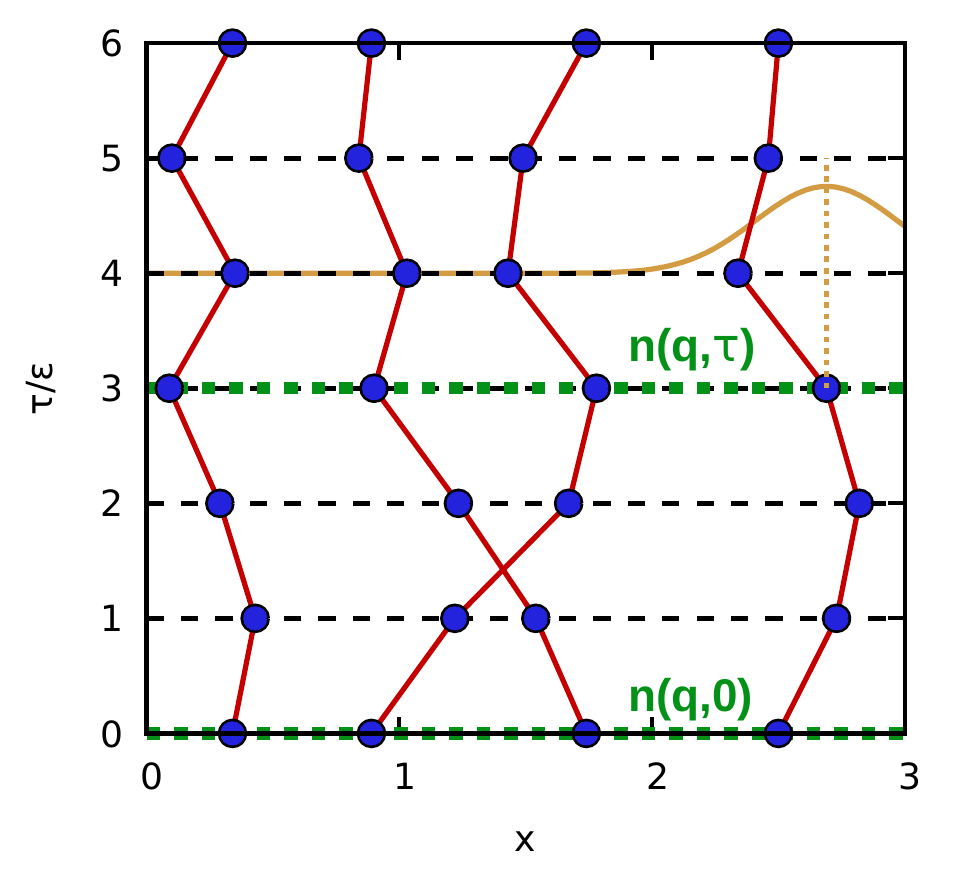}
    \caption{Schematic picture of the imaginary-time evolution of 4 particles in one dimension. The spatial dimension is depicted here as the $x$-axis while the imaginary time axis $\tau$ is shown here in units of expansion steps $\varepsilon$. Each particle path is closed with respect to $x(0) = x(\beta) = x(6 \epsilon)$. Note that the second and third particles are involved in a so-called permutation cycle~\cite{Dornheim_permutation_cycles} and form a single intertwined path with the same $2\beta$-periodic boundary conditions. The green dashed lines represent densities at $\tau=0$ and $\tau = 3$. The evaluation at each time step of the density enables the computation of the imaginary-time correlation function. The yellow gaussian curve represents the kinetic configuration weight of the particle, which in turn governs the diffusion process along the imaginary-time axis. This diffusion process is responsible for the decay of the ITCF from $\tau \in [0, \beta/2]$. Figure taken from Ref.~\cite{Dornheim2022Physical} with the permission of the authors.}
    \label{fig:ITCF_concept}
\end{figure}

It is well known from statistical mechanics that one can obtain any ensemble average from the partition function. 
In principle, \Cref{eq:Part_function} enables us to calculate all equilibrium ensemble averages of the system from a trace over the density operator  \Cref{eq:Th_avg}.
This includes the density--density corelation function of the system. However, a particularly useful feature of the imaginary-time formalism reveals itself if one expands the partition function into $P$ imaginary-time slices, which allows an accurate evaluation even for complex many-body systems. For simplicity, we focus here on the case of distinguishable particles. The observations remain the same in the case of indistinguishable particles. Let $\mathbf{R} = (\mathbf{r}_1, \mathbf{r}_2, \dots, \mathbf{r}_N)^{T}$ be the shorthand notation to include all spatial integration coordinates of the $N$-particles.  The corresponding partition function can be expanded by inserting $P-1$ spatial unities into an alternative form
\begin{align}
    Z &= \int \, d\mathbf{R} \ d\mathbf{R}_1 \dots d\mathbf{R}_{P-1} \bra{\mathbf{R}} e^{-\varepsilon \hat{H} }\ket{\mathbf{R}_1}  \nonumber \\ &\times \bra{\mathbf{R}_1} e^{-\varepsilon \hat{H} }\ket{\mathbf{R}_2} \dots
    \bra{\mathbf{R}_{P-1}} e^{-\varepsilon \hat{H} }\ket{\mathbf{R}} \nonumber \\
    &= \int \, d\mathbf{R} \ d\mathbf{R}_1 \ \dots d\mathbf{R}_{P-1} \prod_{i=0}^{P-1} \rho(\mathbf{R}_i,\mathbf{R}_{i+1};\varepsilon),\label{eq:Z_with_paths}
\end{align}%
where $\varepsilon = \beta  / P = (P k_B T)^{-1}$ and $P$ denotes the number of the discretizations steps of the imaginary-time axis. This expansion is essentially a Wick rotation of the Feynman path-integral \cite{kleinert2009path}. To reconstruct the dynamic properties of the equilibrium system, one can in theory use an analytic continuation\cite{chuna2025dualformulationmaximumentropy}, which is a challenging practical task. \Cref{eq:Z_with_paths} describes the paths of multiple particles in imaginary-time that are connected by the high-temperature density-matrix elements $\rho(\mathbf{R}_i,\mathbf{R}_{i+1};\varepsilon)$. A conceptual picture of this is shown in \Cref{fig:ITCF_concept}.

This is the starting point of how to derive highly efficient sampling schemes for the partition function such as path-integral Monte Carlo (PIMC) \cite{boninsegni1,boninsegni2}. % One must sample closed paths in imaginary time $\tau \in [0,\beta]$  to obtain exact ensemble averages of the system.

\subsection{Imaginary-time correlation functions}
\label{ssec:ITCF}
An important avenue opened by \Cref{eq:Z_with_paths} is the possibility to calculate the equilibrium density fluctuations of the system from the thermal average defined by \Cref{eq:Th_avg}
\begin{equation}
\label{eq:ITCF_PIMC}
F(\mathbf{k},\tau) = \braket{\hat{n}(\mathbf{k},0)\hat{n}(-\mathbf{k},\tau)}, 
\end{equation}
which describes the density diffusion of particles within imaginary time where $\hat{n}(\mathbf{k},\tau)$ refers to the imaginary-time density particle operator. This function is called the imaginary-time correlation function (ITCF). It is defined analogously to \Cref{eq:density_rt}. %Since we only performed a Wick rotation of the real-time propagator, \Cref{eq:ITCF_PIMC} must contain the same physical information as the real-time density--density intermediate scattering function \Cref{eq:intermediate}.
After the Wick rotation of the closed-path expressions, one can obtain again the equilibrium real-time dynamic canonical averages using analytic continuations, which may be challenging in practice. In an experiment, the density fluctuations of the system can be probed by scattering experiments under the assumption that the probe beam is sufficiently weak to not significantly influence the sample equilibrium distribution. A scattering experiment will measure not the real-time intermediate scattering function \Cref{eq:intermediate}, but the dynamic structure factor \Cref{eq:DSF_FT}. The connection between the DSF and the ITCF is given by performing the inverse Fourier transformation in \Cref{eq:DSF_FT} and a Wick rotation that results in a two-sided Laplace transform
\begin{equation}\label{eq:analytic_cont}
    F(\mathbf{k},\tau) = \int_{-\infty}^{+\infty} d\omega \ S(\mathbf{k},\omega) e^{-\tau \omega} = \mc{L}[S(\mbf{k},\omega)]. 
\end{equation}
The full calculation to arrive at this result requires some technical steps and results from complex analysis that we do not reiterate for simplicity. This equation tells us that in order to compare the results from an ensemble averaged simulation with the real-time measurements, we have to perform an inverse Laplace transformation~\cite{JARRELL1996133}.

Inversion of such class of transformations are unfortunately a notoriously difficult, exponentially ill-posed problem that, so far, has been solved rigorously only in special cases~\cite{DornheimPRL2018}. Recent promising advances have shown remarkable results such as a second roton feature of the uniform electron gas \cite{ChunaSecondRoton,chuna2025dualformulationmaximumentropy,4d4b-kgtk}, but a general solution to the analytic continuation problem currently remains out of reach.
Elevated from its original context, \Cref{eq:analytic_cont} has been reinterpreted in Ref.~\cite{Dornheim_T_2022} to circumvent this problem entirely. In equilibrium, analytic continuation provides a formal route between imaginary-time and real-frequency representations, so that, under the usual analytic assumptions, either representation determines the other. In particular, the imaginary-time correlation function can be obtained from a measurement of $S(\mathbf{k},\omega)$ via the corresponding integral transform. %These considerations bring us to the recognition that the ITCF must also have a degree of sensitivity to the system parameters and subsequent observables that can be obtained from the DSF.
Thus, the ITCF as well as the DSF is sensitive to the system parameters and may be used to infer system parameters or otherwise gain insight into the system.
As we discussed in the previous subsection, experiments do not probe the DSF directly, but rather the convolution between the SIF and the DSF. Therefore, in order to access the temperature information contained in the spectrum one has to also obtain a deconvolved ITCF. In contrast to Fourier space, deconvolution in the Laplace domain is more stable with respect to the experimental noise. One can use the analogous identity to the Fourier deconvolution in Laplace space

\begin{equation} \label{eq:deconv_lap}
    \mathcal{L}[S(\mathbf{k},\omega)](\tau) = \frac{\mathcal{L}[S(\mathbf{k},\omega) \circledast R(\omega)]}{\mathcal{L}[R(\omega)]},
\end{equation}
where the functional operator $\mathcal{L}[\cdot]$ refers to the two sided Laplace transformation, defined by
\begin{equation}
    \mathcal{L}[h(\omega)](\tau) = \int_{-\infty}^{+\infty} \text{d}\omega \ h(\omega) \ e^{-\tau \omega}.
\end{equation}
\Cref{eq:deconv_lap} reveals that in order to infer the full information in XRTS, one must have precise knowledge of the SIF $R(\omega)$. This problem persists also in forward modeling, where the convolution of a theoretical DSF model and $R(\omega)$ is fit to the measured intensity signal. Unfortunately, in certain experimental setups, e.g., at the NIF, one does not have direct access to $R(\omega)$ during the experiment and it has to be modeled~\cite{Gawne_JAP_2024,gawne2025heartnewxraytracing} or characterized in a different campaign~\cite{MacDonald_POP_2022}. In the following section we will go through a practical example on how to utilize the ITCF for inferring the temperature.

\begin{figure}
    \centering
    \includegraphics[width=0.485\textwidth]{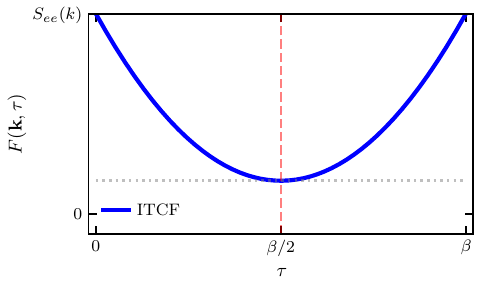}
    \caption{Schematic curve shape of an ITCF shown in blue. The ITCF is a well defined correlation function on the imaginary-time axis $\tau \in [0, \beta]$. The minimum at $\beta/2$ is shown as the red dashed line. An important point of a properly normalized ITCF is at its origin value at $\tau=0$, which coincides with the total static structure factor of the system.}
    \label{fig:itcf_ex}
\end{figure}

To enhance the readers general understanding of  the ITCF as a correlation function object, we show a schematic example of the typical curve shape of an ITCF in \Cref{fig:itcf_ex}. The blue curve depicts the general shape of an ITCF, which is symmetric around $\beta/2$ in thermal equilibrium by detailed balance [\Cref{eq:detailed_balance_dsf}].  This can be shown by applying \Cref{eq:detailed_balance_dsf} to \Cref{eq:analytic_cont}:

\begin{align}
    F(\mathbf{k},\tau) &= \int_{-\infty}^{\infty} \ d\omega \ e^{-\tau \omega} \ S(\mathbf{k},\omega) \nonumber \\
    &=  \int_{0}^{\infty} \ d\omega \ \bigg( \ e^{-\tau \omega} S(\mathbf{k},\omega) + S(\mathbf{k},-\omega) e^{\tau \omega}\bigg) \nonumber \\
    &= \int_{0}^{\infty} \ d\omega \ S(\mathbf{k},\omega) \left( e^{-\tau \omega} + e^{-\omega (\beta - \tau)} \right) \nonumber \\
    &= F(\mathbf{k},\beta - \tau).\label{eq:ITCF_Symmetry}
\end{align}
Therefore, if one can compute the exact ITCF from a measurement, the temperature immediately follows from the symmetry point of the ITCF using

\begin{equation}
    k_BT = \frac{1}{2 \tau_\text{min}},
\end{equation}
where $\tau_\text{min}$ refers to the location of the minimum of the ITCF.
% Another special point of the ITCF is its value at $\tau=0$, that requires, in the case of a performed deconvolution, a proper normalization that obeys the f-sum rule as will be discussed in \Cref{ssec:fsum}.  If we look at \Cref{eq:analytic_cont} , it is easy  to see that the value at $\tau=0$ coincides with the total static structure factor (SSF) of the system
% \begin{equation}
%     F(\mathbf{k},0) = \int_{-\infty}^{+\infty} d\omega  \ S(\mathbf{k},\omega) = S(\mathbf{k}). 
% \end{equation}

Additionally, the value of the ITCF at $\tau=0$ corresponds to the total static structure factor (SSF) $S(\mathbf{k})$ of the system:
\begin{equation}
    F(\mathbf{k},0) = \int_{-\infty}^{+\infty} d\omega  \ S(\mathbf{k},\omega) = S(\mathbf{k}). 
\end{equation}
This fact is leveraged in \Cref{ssec:fsum} to infer the proper normalization of the ITCF from the (unnormalized) Laplace transform of the experimentally measured DSF.

We stress here again that the deconvolved ITCF from \Cref{eq:deconv_lap} does not immediately deliver the  SSF but must be normalized to adhere to the f-sum rule. Therefore, the ITCF is a useful object that delivers insights into the equilibrium properties of system and can be computed immediately from experimental measurements for the case of a precisely known SIF. 

\section{Methods\label{sec:results}}
\begin{figure}
    \centering
    \includegraphics[width=0.485\textwidth]{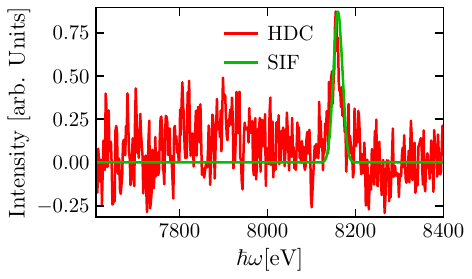}
    \caption{XRTS spectrum in arbitrary units of heated diamond as reported in Ref.~\cite{Martin2025}. The green curve shows the corresponding source-and-instrument function (SIF) for the reported shot. The spectrum was recorded with a beam energy of 8160 eV at a scattering angle of 170$^{\circ}$.}
    \label{fig:Spectrum_raw}
\end{figure}

We start with the spectrum recorded in the experimental campaign reported by Martin et al.~\cite{Martin2025} for diamond heated by a short-pulse laser. The spectrum is shown as the red curve in \Cref{fig:Spectrum_raw} and the SIF is depicted by the green curve normalized to the peak height of the measurement. We aligned the SIF peak at the reported measurement energy of 8160 eV \cite{Martin2025}. The reported inferred temperature for this particular data set was T=$61^{+27}_{-19}$ eV based on a Markov-Chain Monte Carlo analysis using a standard Chihara model assuming a charge state of Z=$4$. The spectrum was recorded at LCLS using a spectrometer placed at $170^{\circ}$ resulting in a wave-vector of $k = 8.24 \, \AA^{-1}$ and therefore in a non-collective scattering geometry. 

\subsection{Temperature from the ITCF}
\label{ssec:temp_xfel}
\begin{figure*}
    \centering
    \includegraphics[width=0.9\textwidth]{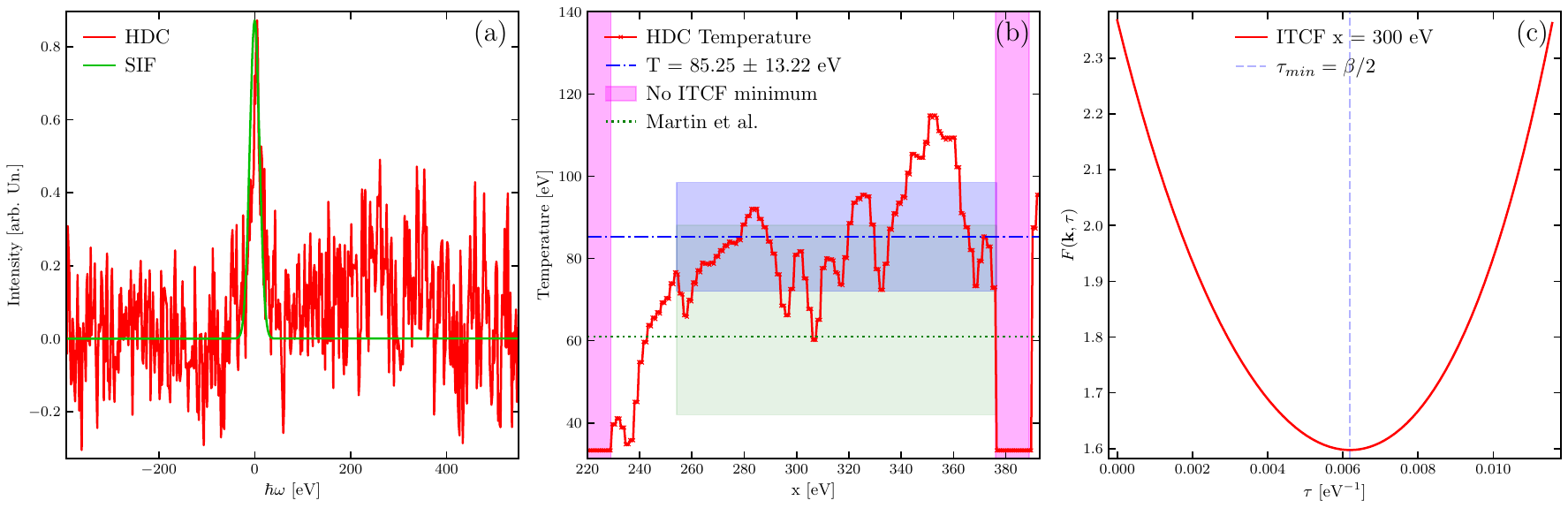}
    \caption{Temperature analysis of the heated diamond spectrum reported in Ref.~\cite{Martin2025}. Panel (a) depicts the reported spectrum (red curve) and SIF (green curve) aligned to $\omega=0$ eV and with the energy axis multiplied by $-1$ to correctly align the down- and upshifted energies on the positive and negative axis, respectively.  Panel (b) shows the inferred temperature from the ITCF as a function of the truncation. Due to the low SNR of the signal, the deconvolution is only able to resolve detailed balance for truncations between $229$ and $377$ eV. The average inferred temperature within this range is given by $\bar{T} = 85.23 \pm 13.22$ eV and shown as the blue dash-dotted line and the corresponding systematic uncertainties as the blue shaded band. The green dashed line depicts the inferred temperature in Ref.~\cite{Martin2025} of $T=61^{+27}_{-19}$ eV, where the green shaded band depicts the uncertainties. The sudden drop in the inferred temperature between $376$ and $389$ eV is attributed to the high noise in this interval, which prevents the resolution of detailed balance. Panel (a) depicts the corresponding areas also by the magenta color. Panel (c) shows the calculated ITCF for a truncation of $300$ eV and the vertical blue dashed line represents the location of the minimum.}
    \label{fig:t_analysis}
\end{figure*}

The following considerations have to be made to obtain the Laplace deconvolved version of the spectrum. Experiments measure the density fluctuations of the plasma using a beam at a certain probe energy $\omega_i$. The nature of scattering experiments dictates that the detected energies of the scattered photons $ \omega_s$ are connected to the physical frequency axis by
\begin{equation}
    \omega_s =  \omega_i - \omega. 
\end{equation}
All theories describing the DSFs are defined with respect to $\omega$, which represents the internal transition between states, while the scattering signal is recorded with respect to the scattered radiation frequency $\omega_s$. In this work, we use the convention to refer to the downshifted part of the spectrum for $\omega > 0$ and the upshifted part of the spectrum as $\omega  < 0$. This convention is motivated by the fact that an observed photon with frequency $\omega_s$ get red-shifted for the case of $\omega > 0$ and blue-shifted for $\omega < 0$ compared to the incoming photon of frequency $\omega_i$.
Therefore, one has to align the measurement spectrum to $\omega=0$ by subtracting the probe beam energy and invert the energy axis, such that the down-shifted part of the spectrum and the up-shifted part of the spectrum are aligned at $\omega>0$ and $\omega < 0$, respectively. The same must be done for the SIF.  \Cref{fig:t_analysis} (a) depicts how the spectrum and SIF must be aligned. We then calculate the deconvolved ITCF and obtain a temperature. A vital practical detail for this calculation is that formally the deconvolution in Laplace space involves an infinite integral as shown in \Cref{eq:deconv_lap}. However, detector setups in practice only allow to cover a finite amount of photon energies. Consequently, we have to truncate the Laplace deconvolution at a finite value $x$ 
\begin{equation} \label{eq:Laplace_deconv_trunc}
    \mathcal{L}_x[S(\mathbf{k},\omega)] = \frac{\mathcal{L}_x[S(\mathbf{k},\omega) \circledast R(\omega)]}{\mathcal{L}_{x'}[R(\omega)]},
\end{equation}
with the truncated two-sided Laplace transform defined as 
\begin{equation}
    \mathcal{L}_x[S(\omega)](\tau) = \int_{-x}^{x} d\omega \ S(\omega) e^{-\omega\tau}.
\end{equation}
It is important to note here that the Laplace transformation of $R(\omega)$ can be converged independent of the chosen truncation limit $x$ of the experimental spectrum. We therefore denote the truncation of the $R(\omega)$ Laplacian with $x'$. 
For the deconvolution, we use a model SIF as in Ref.~\cite{Martin2025} of the form 
\begin{equation}\label{eq:SIF_willo}
    R(\omega) = \sqrt{\frac{\ln{2}}{\pi}} \frac{1}{\alpha}\exp\left[\frac{-(\omega - \omega_i)^2}{\alpha^2} \ln{2}\right], 
\end{equation}
with $\omega_i = 8160 eV$ and $\alpha=12$. The result of the deconvolution is shown in panel (c) of \Cref{fig:t_analysis}. The panel shows the ITCF of the data from the previous panel for a cutoff energy of $x=300$ eV as a function of the imaginary time $\tau$. The computed ITCF curve exhibits a clear symmetry around its minimum. This is a direct consequence of detailed balance in the DSF \Cref{eq:detailed_balance_dsf}. Equation (\ref{eq:ITCF_Symmetry}) shows that detailed-balance directly implies that the ITCF must be symmetric at $\tau=\beta / 2$ of the system temperature and also has a minimum there. It therefore enables us to directly infer the presumed temperature of the system if we have a recorded XRTS intensity and a sufficient truncation interval. However, to obtain an accurate temperature, one cannot truncate the Laplace deconvolution arbitrarily, but must check which truncation is sufficient until the temperature remains roughly constant. 

%%%%%%%
% New interpretation of the experiment
%%%%%%
Panel (b) of \Cref{fig:t_analysis} shows the inferred temperature from \Cref{eq:Laplace_deconv_trunc} as a function of the truncation parameter $x$. Because of the low SNR, the deconvolution reliably resolves detailed balance only for truncations in the range $229$–$377$ eV. The inferred temperature begins to oscillate about a mean for $x>254$ eV. We therefore estimate the temperature and its systematic uncertainty by taking the average and standard deviation of the inferred values for $254 \le x \le 377$ eV, yielding $\bar{T}=85 \pm 13\ \text{eV}$ (standard deviation). In Fig. \Cref{fig:t_analysis}(b), the red curve shows the temperature estimated at each truncation limit $x$; the blue dash-dotted line and shaded region depict $\bar{T}$ and its systematic uncertainty. For comparison, our estimate is consistent with the analysis of Ref.~\cite{Martin_POP_2025} (green dotted line with shaded uncertainties), which reported $T=61^{+27}_{-19}\ \text{eV}$. A prominent feature in panel (b) is a sudden decrease in the inferred temperature for $x\in[376,389]$ eV. In this interval the noise level prevents a reliable computation of the ITCF; the ITCF monotonically decreases, and no minimum is found. Consequently, temperatures inferred with truncation values in this range are unreliable. Beyond this interval the curve suggests some recovery of detailed balance, but the higher noise level in the data leads us to exclude those truncations from the average. Because the upshifted feature extends only up to an energy transfer of $-393$ eV, the detector range precludes larger truncations. Accordingly, while our result agrees with Ref.~\cite{Martin_POP_2025}, it should be regarded as an estimate. We determine that a higher SNR is required for a conclusive result. Nevertheless, the partial recovery of detailed balance and consistency with prior results indicate that the Laplace transform is relatively robust to noise. Consequently, the ITCF approach still yields meaningful estimates and provides a straightforward way to quantify uncertainty of temperatures from data with high noise or limited photon-counting statistics.

This exercise shows that the computation of temperature using the ITCF is not only simple, but also robust against noise. A corollary of this method is to reconstruct the temperature with a high degree of certainty, one must know the SIF of the measurement setup as precisely as possible as it is the only assumed quantity in the calculation. Furthermore, the symmetry in the ITCF directly reflects how well detailed-balance and therefore local thermal equilibrium is preserved in the measurement data. The ITCF is not only useful to infer the temperature of the system, but can also be used to gain further physical insights. 
A detailed list of the steps that have to be taken for the temperature extraction is shown in \Cref{sec:workflow}. 

However, one important note here is that the ITCF calculated with a cutoff, that does not cover the whole spectral range of the XRTS spectrum, is not the physical ITCF of the system. For instance, the ITCF shown in \Cref{fig:t_analysis} (c) does not reflect the real ITCF of the system because the truncation of the deconvolution is not sufficiently large to cover the whole recorded spectrum. The extracted ITCF rather is a correlation function that obeys the same detailed balance relationship as the full ITCF, which still enables estimate of a temperature. Due to the low SNR in \Cref{fig:t_analysis}, the subsequent methods cannot be applied directly, and we therefore will proceed using a synthetic spectrum.

\subsection{Normalization of the ITCF and SSF from the f-sum rule}\label{ssec:fsum}
We now compute the absolute normalization of the inferred ITCF, following the approach of Ref.~\cite{dornheim2023xray}. The goal of this exercise is to obtain an absolute normalization from the f-sum rule without requiring a fully deconvolved DSF.  Issues arise due to \Cref{eq:deconv_lap} only determining the ITCF up to a normalization factor $A$ defined as
\begin{align}
    A F(\mathbf{k},\tau) &= F_\text{exp.}(\mathbf{k},\tau) = \frac{\lap{S(\mathbf{k},\omega)\circledast R(\omega)}}{\lap{R(\omega)}}.
\end{align}
$F_\text{exp.}$ refers to the experimentally inferred ITCF, which differs in normalization from the ITCF of \Cref{eq:ITCF_PIMC} by only the normalization constant $A$.

We define the frequency moment of degree $\alpha$ of the DSF by
\begin{equation}
    M_\alpha^S = \int_{-\infty}^{+\infty} S(\mbf{k},\omega) \omega^{\alpha} \dom .
\end{equation}
The f-sum rule \cite{quantum_theory} states that the first frequency moment of the DSF given by
\begin{align}
    M^S_1 &= \int_{-\infty}^{+\infty} d\omega \ S(\mathbf{k},\omega) \omega \nonumber \\
    &= -\frac{\partial}{\partial \tau} \frac{\lap{S(\mathbf{k},\omega) \circledast R(\omega)}}{\lap{R(\omega)}}\bigg|_{\tau=0}= \frac{\mbf{k}^2}{2 m_e},\label{eq:f-sum}
\end{align}
must be obeyed due to particle number conservation.
Since the f-sum rule has to be obeyed for all $\mathbf{k}$, we can define an absolute normalization of the ITCF that ensures the f-sum rule holds from a measurement at a single scattering angle. This is done by connecting the first frequency moment of the DSF to the ITCF given by
\begin{equation}\label{eq:normalisation}
    A =  - \frac{2m_e}{\mathbf{k}^2}\frac{\partial}{\partial \tau} \frac{\mathcal{L}[I(\mathbf{k},\omega)]}{\mathcal{L}[R(\omega)]} \bigg|_{\tau = 0}.
\end{equation}
In practice, this requires that the spectra have been measured over an energy range such that the limits of the Laplace integral are effectively infinite, and with sufficient signal to distinguish the data from the background. 

\begin{figure}
    \centering
    \includegraphics[width=0.485\textwidth]{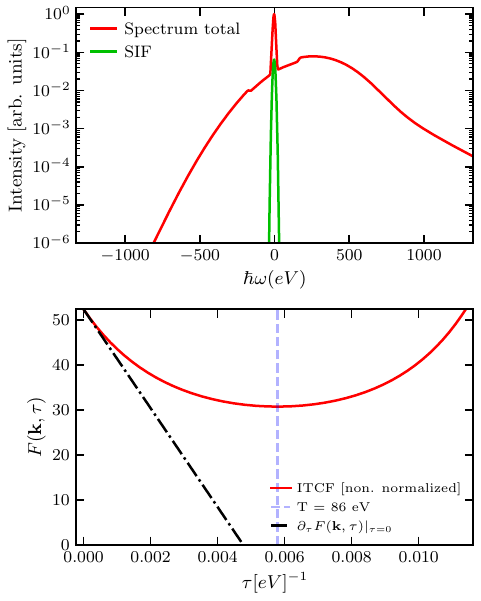}
    \caption{Synthetic XRTS spectrum of $C^{4.5+}$ at $T=86$ eV, $\rho=5$ g\unitspace cm$^{-3}$, $\Theta = 170^{\circ}$ calculated using the Chihara decomposition \cite{Gregori_PRE_2003}. The lower panel depicts the corresponding ITCF integrated within the interval $\omega \in [-1330, 1330]$ eV. The black dashed-dotted line depicts the slope of the ITCF derivative at $\tau=0$ needed to obtain the normalization constant from the f-sum rule.}
    \label{fig:f_sum_syn}
\end{figure}

To demonstrate how to obtain the absolute normalization for a spectrum, we computed a spectrum using the Chihara decomposition \cite{Chihara_1987,Gregori_PRE_2003} for $C^{4.5+}$ at $T=86$ eV, $\rho=5$ gcm$^{-3}$ and $\Theta = 170^{\circ}$ in the top panel of \Cref{fig:f_sum_syn}. The bottom panel depicts the corresponding deconvolved ITCF where we chose a truncation of 1330 eV as any higher cutoff will result in issues with resolution of the down-shifted feature due to numerical instabilities of floating point operations. The cutoff sufficiently resolves both the upshifted and downshifted feature to calculate an accurate ITCF, shown in the lower panel. The black dashed-dotted line depicts the slope of the $\tau = 0$ derivative using a forward numerical derivative, that we require for the normalization constant. 
In Hartree atomic units, we require the square of the scattering wave-vector that we can approximate if the incoming and outgoing frequencies $\omega_{i}/\omega_s \approx 1$ \cite{Thermometry2023} 
\begin{equation}
    k \equiv |\mathbf{k}| \approx 2 k_0 \sin(\Theta/2).
\end{equation}
In our case, we have $k_0 = \omega /c = E_0/\hbar c = 4.14 \times  10^{10}$ m$^{-1}$. Thus the wave-number is given by $k = 8.24 \times 10^{10}$ m$^{-1}$ $= 4.36 a_B^{-1}$. 

The normalization of the ITCF opens a pathway to many practical applications\cite{dornheim2024modelfreerayleighweightxray,schwalbe2025staticlineardensityresponse}, such as the extraction of the density response function and inference of the mass density.  Refs.~\cite{Dornheim2025-sv,schwalbe2025staticlineardensityresponse} also introduce another practical way  to obtain both the normalization and the electron-electron static structure factor (SSF) $S_{ee}(k)$. Comparing the definition of $S_{ee}(k)$ with $F(k,\tau)$
\begin{align}
    S_{ee}(k) &= \int_{-\infty}^{+\infty} d\omega \  S_{ee}(k,\omega) \nonumber \\
    &= M_0^{S}(k) = F(k,0),\label{eq:SSF}
\end{align}
implies that the absolute normalization of the ITCF immediately delivers the static electronic structure factor. 

\begin{figure}
    \centering
    \includegraphics[width=0.485\textwidth]{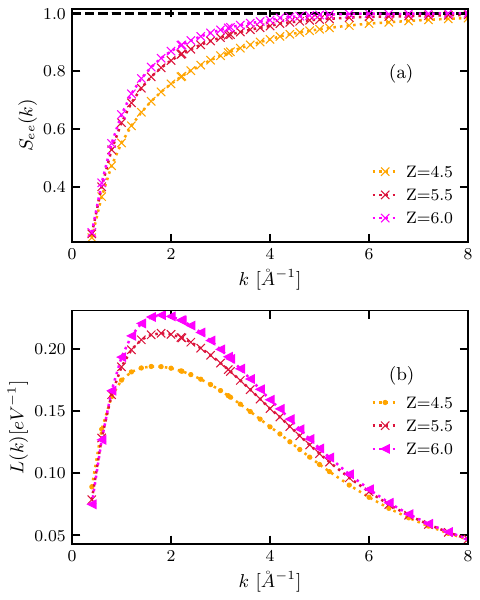}
    \caption{Panel (a) depicts the static structure factor of carbon as a function of the wave-vector for several charge states at $\rho=5 \ \mathrm{g\unitspace cm^{-3}}$ and $T=86$ eV computed from the Chihara decomposition. Panel (b) depicts the resulting area under the ITCF curve from the same conditions.}
    \label{fig:See_L}
\end{figure}

We show the SSF for this example in C at several ionization degrees in \Cref{fig:See_L} panel (a). $S_{ee}(k)$ shows the expected behavior of converging towards unity for large wavenumbers $k$. From a practical perspective, calculating the SSF and the normalization from \Cref{eq:normalisation,eq:SSF} involves an integration close to $\tau=0$. Consequently, computing the SSF and the normalization is more robust with respect to the SNR of the upshifted part of the spectrum. One of the central results in Ref.~\cite{dornheim2023xray} is that the slope in the ITCF can be too steep to allow for practical calculations using a simple forward numerical derivative. A more practical and accurate approach is to use a polynomial fit in order to compute the $\tau=0$ derivative~\cite{Dornheim_moments_2023}. The convergence behavior of the static structure factor for carbon is shown as an example for the case of $C^{4.5+}$ at $k=8 \AA$ in \Cref{fig:Convergence}. 
%%%% Insert Carbon example here for S_ee and A
\subsection{Rayleigh weight from the static structure factor}\label{ssec:Wr}
We present how to practically extract the Rayleigh weight $W_R$ from measurements \cite{Dornheim2025-sv,dornheim2024modelfreerayleighweightxray}. $W_R$ itself is a measure of the total electronic localisation around the nuclei in a plasma. For instance, in the Chiahara decomposition, it reflects the amount of elastic scattering from electrons localised around the ions
\begin{equation}
    S^\text{Ch.}(k,\omega) = S_\text{inel.}(k,\omega)  + W_R^\text{Ch.}(k) \delta(\omega). 
\end{equation}
Furthermore $W_R(\mathbf{k})$ is generally defined as \cite{Vorberger2015}
\begin{equation}
    W_R(k) = \frac{S_{eI}^2(k)}{S_{II}(k)},
\end{equation}
where $S_{eI}(k)$ refers to the electron-ion static structure factor and $S_\text{II}(k)$ refers to the ion-ion static structure factor. In the aforementioned Chihara decomposition the Rayleigh weight is modeled as \cite{Wunsch2011}
\begin{equation}
    W_R^\text{Ch.}(k)= \sum_{a,b} \sqrt{n_a n_b}\,\lvert f_a(k) + q_a(k)\rvert\,\lvert f_b(k) + q_b(k)\rvert\, S_{ab}(k),
\end{equation}
where the indices $a,b$ run over all ion species in the plasma, $n_a$ and $n_b$ denote the partial number densities of the respective ion species, $f_a(k)$ and $f_b(k)$ are the corresponding ionic form factors, and $q_a(k)$ and $q_b(k)$ represent the associated screening clouds. The ionic form factor of species $a$ is defined as the Fourier transform of the bound-electron density around an ion of species $a$. $S_{ab}(k)$ denotes the partial static structure factor describing correlations between ion species $a$ and $b$. In Chihara-type approaches, ions are usually treated classically, and $S_{ab}(k)$ is often computed \cite{Wuensch2008} using hyper-netted-chain methods or obtained from DFT-MD simulations \cite{Wunsch2009} .

\begin{figure}
    \centering
    \includegraphics[width=0.485\textwidth]{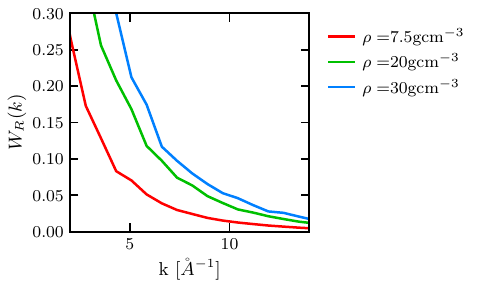}
    \caption{Rayleigh weight of warm dense beryllium as a function of scattering wave-vector. The curves show a corresponding calculation of the Rayleigh weight from spectral partitioning DFT \cite{Sadigh2023} using the PBE functional~\cite{PBE} at $7.5, 20$ and $30$ g/cm$^{3}$ for a temperature of 150 eV.}
    \label{fig:WR}
\end{figure}

A common observable that can be extracted from precise knowledge of the SIF is the ratio of elastic to inelastic scattering, obtained by decomposing the spectrum into its elastic and inelastic components. For this exercise, one only has to obtain a fit of the experimental spectrum that accurately can capture the elastic feature and compute 
\begin{equation}\label{eq:el_inel_rat}
    r = \frac{\int_{-\infty}^{+\infty} I_\text{el}(k,\omega) d\omega}{\int_{-\infty}^{+\infty} I_\text{inel}(k,\omega) d\omega} =  \frac{W_R(k)}{S_{ee}(k) - W_R(k)}.
\end{equation}
Due to the knowledge of $S_{ee}(k)$ from \Cref{eq:SSF,eq:normalisation}, we can infer the Rayleigh weight by 
\begin{equation}\label{eq:Wr}
    W_R(k) = \frac{S_{ee}(k)}{1 + r^{-1}(k)},
\end{equation}
from known quantities that can be computed cheaply without any model assumptions. This approach has opened up a way to successfully infer densities from XRTS experiments at the National Ignition Facility (NIF) for beryllium capsule implosions \cite{Tilo_Nature_2023,Dornheim2025-sv}. Additionally, the knowledge of $W_R$ opens avenues to discriminate between model assumptions that are often used in WDM calculations, such as different exchange-correlation functionals. An example is depicted in \Cref{fig:WR}. The figure shows the Rayleigh weight computed from density functional theory using the SPARC code \cite{xu2021sparc,ZHANG2024100649} with spectral partitioning \cite{Sadigh2023} at 150 eV for various wave-numbers at different densities. %This is compared to the experimental data point obtained in Ref.~\cite{Tilo_Nature_2023}, which is shown as the red cross. The curve suggest that the density of the sample must be between 20 \gcc and 30 \gcc. This is in good agreement with the results found in Ref.~\cite{dornheim2024modelfreerayleighweightxray}. 
%%%%%%%% Insert Carbon example here for r and Wr
\subsection{Density from the area under the ITCF} \label{ssec:area}
\begin{figure}
    \centering
    \includegraphics[width=0.485\textwidth]{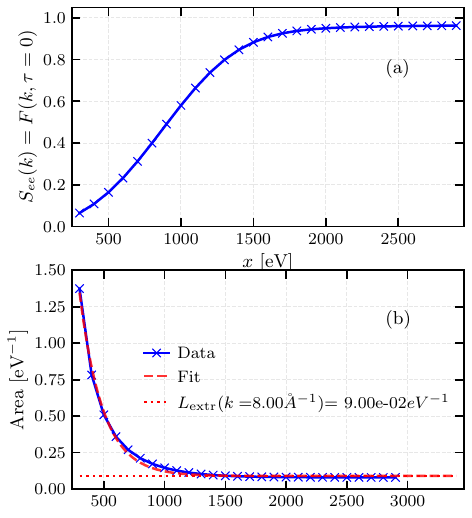}
    \caption{Convergence behavior of the (a) static structure factor and (b) the area under the ITCF curve \Cref{eq:area_itcf} for $C^{4.5+}$ for the conditions specified in \Cref{fig:See_L} as a function of the truncation $x$ as the blue crosses at $k=8\AA$. The red dashed line shows an exponential extrapolation fit to overcome the numerical limitation with the cutoff radius of the area under the ITCF curve. The vertical red dotted lines in each panel depict the asymptotic extrapolation for $x\rightarrow \infty$.
    } 
    \label{fig:Convergence}
\end{figure}

A useful relation is the imaginary-time version of the fluctuation-dissipation theorem \cite{Dornheim2022Physical,schwalbe2025staticlineardensityresponse}
\begin{equation}\label{eq:FDT}
    \chi(k,\omega=0) = -n \underbrace{\int_{0}^{\beta} \ d\tau \ F(k,\tau)}_{=:L(k)},
\end{equation}
where $\chi(k,\omega)$ refers to the linear density response function \cite{quantum_theory,Dornheim_review}. This equation reveals an interesting connection. The area under the ITCF curve $L(k)$ is an experimental observable that can be computed as soon as the normalization \Cref{eq:normalisation} is obtained. The static density response function $\chi(k,0)$ on the left hand side \Cref{eq:FDT} is an accessible quantity that can be calculated even from single-particle theories \cite{BoehmePRE2023,Bohme_PRL_2022,Moldabekov2023}. \Cref{fig:See_L} panel (b) depicts the behavior for the area under the ITCF for carbon computed using the Chihara decomposition. As shown in Schwalbe et al.~\cite{schwalbe2025staticlineardensityresponse}, this calculation can then be utilized in comparison with different theories capable of calculating $\chi(k,0)$ to infer densities from experimental measurements. 
One important prerequisite for this comparison is that the ITCF is very well converged with respect to the truncation of \Cref{eq:deconv_lap}. Panel (b) of \Cref{fig:Convergence} depicts the convergence of $L(k)$ at $k = 8 \AA^{-1}$ with respect to the truncation from synthetic Chihara spectra of $C^{4.5+}$. One can here use an exponential fit 
\begin{equation}
    f(x) = a + b e^{cx},
\end{equation}
to overcome numerical or practical limitations, which is depicted as the red dashed curve. The extrapolated area under the ITCF curve is shown as the dotted red lines in the figure.
As a practical method, Ref.~\cite{schwalbe2025staticlineardensityresponse} introduces a two-fold convergence approach on how to practically obtain both the normalization and the SSF by introducing a double truncated Laplace deconvolution
\begin{equation}
\label{eq:area_itcf}
    \mathcal{L}_{x,y}[S(k,\omega)] = \int_{x}^{y} \ d\omega \ S(k,\omega)  \ e^{-\tau \omega}.
\end{equation}
This can now be used in the deconvolution integral 
\begin{equation}
    \mathcal{L}[S(k,\omega)] = \frac{\mathcal{L}_{x,y}[S(k,\omega)\circledast R(\omega)]}{\mathcal{L}_{x'}[R(\omega)]},
\end{equation}
where one must check the double-sided convergence of the deconvolution with respect to $x$ and $y$. 
By employing this two-fold convergence, the ITCF can be captured with sufficient fidelity, and its normalization can be evaluated with the accuracy required to infer the density from \Cref{eq:FDT}. 

In conjunction with \Cref{ssec:Wr}, this enables two independent pathways for determining densities from XRTS spectra: one based on comparison with theoretical approaches capable of computing the Rayleigh weight, and another relying on methods that provide the static density response function. Accordingly, the recommended practice for density inference from XRTS is to combine both approaches, drawing on distinct models or \textit{ab initio} techniques. This integrated strategy establishes a robust framework for density determination that can be applied in a direct and systematic manner.
%\subsection{Error and noise analysis}
%One of the central results of the recent works \cite{Thermometry2023,Dornheim_T_2022,Gawne2024} is that in order to fully 
%%%%%%%%%%%%%%%%%%%%%%%%%%%%%%%%%%%%%%%%%%%%%%%%%%%%%%%%%
\subsection{Detecting non-equilibrium for multiple scattering angles}\label{ssec:noneq}
As a last step, we are reviewing the most important results from Ref.~\cite{VORBERGER2024129362}. So far all our computations relied on the fact that the DSF fulfills detailed balance as soon as the probed system is in local thermal equilibrium (LTE). The previous observation can be utilized to not only infer the equilibrium properties of a system, but also to detect non-equilibrium conditions once detailed balance is violated. One of the central results in Ref.~\cite{VORBERGER2024129362} is that the violation of the detailed-balance symmetry condition might not be possible to detect if the resolution of the experiment is insufficient, which may be the case for current experimental setups \cite{Bellenbaum_APL_2025}. 

\begin{figure}
    \centering
    \includegraphics[width=0.445\textwidth]{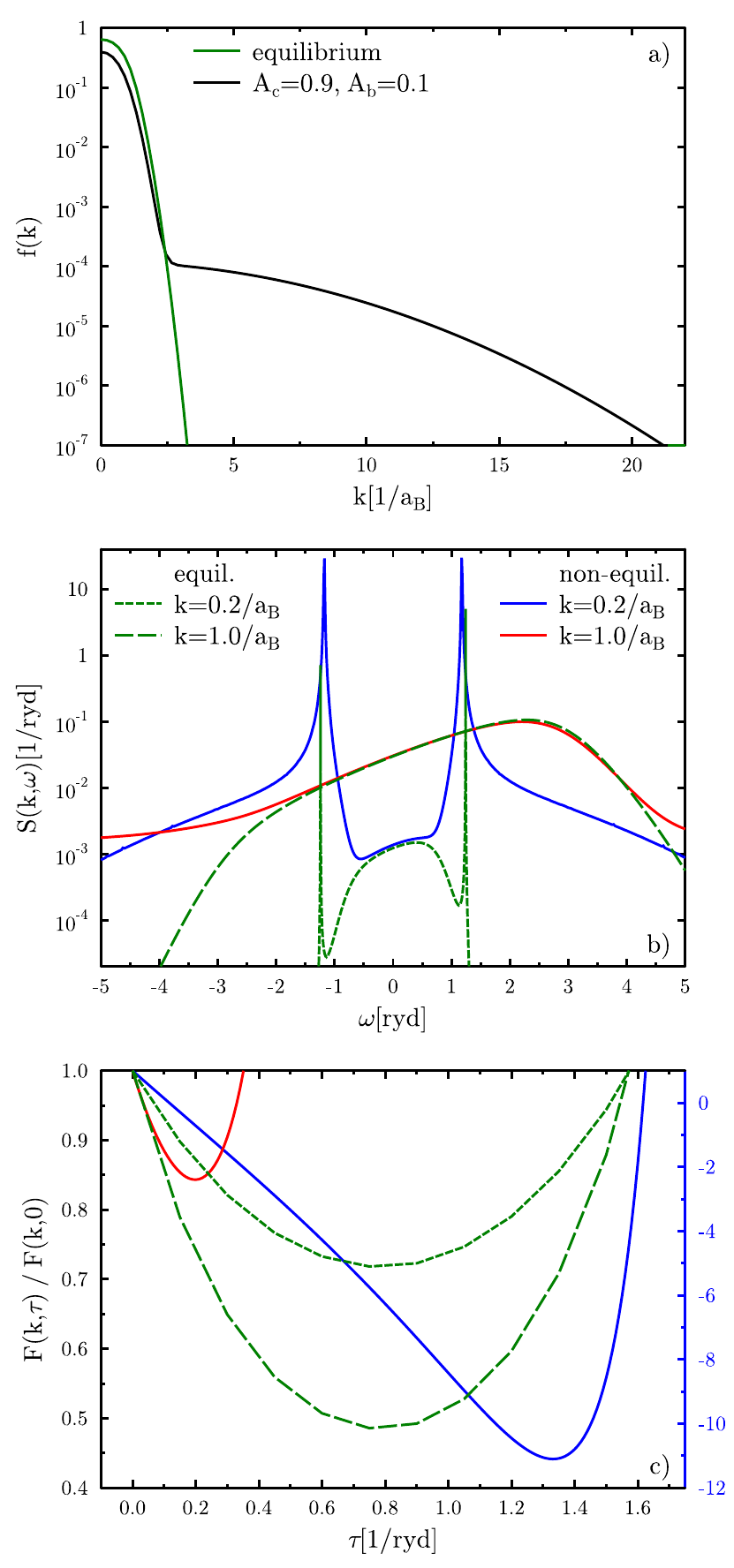}
    \caption{Detection of non-equilibrium as suggested in Ref.~\cite{VORBERGER2024129362}. Panel (a) shows the Wigner function of the uniform electron gas for the equilibrium case denoted as the green curve and the non-equilibrium case as the black curve. The electronic density is as given in Ref.~\cite{VORBERGER2024129362} by $n=1.8 \times 10^{23}\ \mathrm{cm^{-3}}$. The cold electronic bulk of the black curve was assigned a temperature of $T_c=10^{5}$ K. The hot electron bump is located at $p_B = 25/a_B$ and models the non-equilibrium scenraio of the black curve with a temperature of $T_b= 10^4$ K. Panel (b) then denotes the corresponding DSF for two different scattering wave numbers in the collective and single particle regime. The equilibrium DSFs are given by both green curves, while the nonequilibrium DSFs are shown by the blue curve for the collective scattering regime and the red curve for the single-particle regime. The subsequent panel (c) then shows the corresponding ITCFs of the four given curves. The blue curve is plotted on a separate scaled y-axis on the right hand side. 
}
    \label{fig:nonequ}
\end{figure}

Therefore, the work suggests an advantageous pathway to detect non-equilibrium shown in \Cref{fig:nonequ}. 
The first panel depicts the Wigner function of the electron gas for an equilibrium (green curve) and non-equilibrium situation (black curve). For this demonstration, we use the same model for the Wigner function with a hot electronic tail as in Ref.~\cite{VORBERGER2024129362} given by
\begin{align}
    f(k,t) &= A_c \Biggl\{\exp\left[ \beta_c \left( \frac{k^2
    }{2m} - \mu_c \right)\right] + 1 \Biggr\}^{-1} \nonumber \\ 
&+A_h \Biggl\{\exp\left[ \beta_h \left( \frac{k^2
    }{2m} - \mu_h \right)\right] + 1 \Biggr\}^{-1}, \label{eq:Wigner}
%    &+ A_b \frac{n_b \Lambda^3}{2} \Biggl\{ \exp \left[ \beta_b \left( \frac{(k - p_b)^2}{2m}  \right)  \right] \Biggr\}^{-1}, \label{eq:Wigner}
\end{align}
with component amplitude $A_{c,h}$ referring to the cold and hot spectra, $\beta_{c,h}$ the temperatures of the bulk and the hot electronic tail, with their individual chemical potential. This situation is a model for the laser heating of electrons, where a hot electronic bump occurs on a cold fermi-distribution. In general, there are a multitude of possible distribution functions that get created in non-equilibrium situations. \Cref{eq:Wigner} can serve as a model for the transient heating of electrons by a laser pulse. 
%Both the equilibrium and non-equilibrium components can be offset using the $\mu_c$ and $p_b$ parameter, respectively. In the pre-factor of the second term, $\Lambda$ refers to the thermal wavelength of the electrons and $n_b$ to the density of the tail electrons.  
Panel (a) displays the Wigner function \Cref{eq:Wigner} under equilibrium (green) and nonequilibrium (black) conditions. Panel (b) of \Cref{fig:nonequ} presents the corresponding dynamic structure factors (DSFs) for these Wigner functions. For wave-vectors $k=0.2,1.0 \, a_B^{-1}$, the equilibrium DSFs appear as green curves, while the nonequilibrium DSFs are represented by blue curves (collective regime) and red curves (non-collective regime). While the DSFs show very similar features, this changes in panel (c). Here both green curves show the same minimum in the equilibrium case and therefore adhere to detailed balance at the same temperature. However, this situation is completely different for the red curve, which depicts the non-eq.~ITCF at $k=0.2 a_B^{-1}$. While the curve itself seems to be largely symmetric around its minimum, the given minimum is notably not at the temperature of the electronic bulk. The the amount of violation with respect to detailed-balance is not sufficient to detect non-equilibrium with certainty in this case. In contrast, the blue curve, that is plotted on a separate y-axis, clearly is not symmetric around its minimum and the location of the minimum disagrees with the red curve. Therefore, it is vital to observe that a symmetric ITCF at a single wavenumber is only a necessary condition for equilibrium. Instead, to establish a necessary and sufficient test of equilibrium, the ITCF at different scattering wavenumbers must all be symmetric about the same minimum in $\tau$. 
%Ref.~\cite{VORBERGER2024129362} even goes on to show that after equilibration of the system the minima at the wave-numbers must agree. 

This remarkable result enables one to explore the non-equilibrium properties of WDM experimentally from first principles. Current experimental setups such as at the European XFEL could in principle be used to explore things such as equilibration times of pump-probe experiments. Other pump-probe experimental setups could be utilized under the condition that spectra are measured simultaneously from multiple scattering angles. A practical application of detecting equilibrium with two scattering angles on isochorically heated graphite has already been tried in Ref. \cite{Bellenbaum_APL_2025}; however, the detector ranges of the presented experiments may have been insufficient. New experiments with modern detectors could deliver a sufficient detector range to resolve non-equilibrium states.

\section{Workflow and practical details}\label{sec:workflow}

The aim of this section is to provide a comprehensive step-by-step guideline to obtain all observables shown in \Cref{fig:Flowchart} from an experimental measurement. 
\subsubsection{Temperature}
To infer temperature using detailed-balance from an XRTS spectrum, one must take the following practical steps 
\begin{itemize}
    \item[1.] Align both the spectrum and the SIF to $\omega = 0$ by subtracting the probe energy. 
    \item[2.] Multiply the energy axis by -1 to correctly align the up- and downshifted feature with negative and positive frequencies, respectively. 
    \item[3.] Define a grid $\tau \in [0,\tau_{max}]$ of sufficient resolution. If in the following steps no temperature can be detected, one can increase $\tau_{max}$ to see if the calculated ITCF's exhibit a minimum. If $\tau_{max}$ is chosen too large, the resolution of the minimum can suffer from it. Generally, $\tau_{max}$ should be chosen close to $\beta$. 
    \item[4.] Define a truncation range grid $x \in [0,x_{max}]$, and calculate for each grid point the ITCF on the $\tau$ grid from \Cref{eq:Laplace_deconv_trunc}. Note down each $\tau_{min}(x) = \beta(x)/2  = [2k_BT(x)]^{-1}$. The maximum truncation value should be chosen with the frequency-resolved spectrum and observed features, particularly on the down-shifted side,  in mind.
    \item[5.] Check if the temperature converges on the $x$ grid and if the ITCF is symmetric for the converged points. If yes, the converged answer is the inferred temperature.
    \item[6.] If the temperature converges up to a certain point on the $x$ grid and then diverges, check if the ITCF is still symmetric around $\beta/2$ beyond these points. One may adjust the upper bound of the $\tau$-grid and its resolution to see if the divergence still exists.
    \item[7.] If the ITCF starts to become more asymmetric with an increasing integration boundary, one must check if the resolution of the upshifted feature is still sufficient or the experimental conditions are not in equilibrium.
    \item[8.] If the inferred temperature exhibits oscillatory behavior within an interval as in \Cref{fig:t_analysis} (c), detailed balance is only resolved in the measurement up to a certain cutoff. One then can average over the aforementioned interval and compute the standard deviation to estimate the temperature and its systematic uncertainty. 
\end{itemize}

\subsubsection{Normalization and SSF}
Once the temperature has been established, the normalization of the ITCF relies on the having the most accurate ITCF with the largest integration range possible available. 
We summarize to take the following practical steps. We assume the spectrum and SIF have both been properly aligned and are sufficiently well behaved. 

\begin{itemize}
    \item[9.] Perform the Laplace deconvolution \cref{eq:Laplace_deconv_trunc} on the largest possible truncation range, that still sufficiently resolves the downshifted feature. 
    \item[10.] To check if the downshifted feature is properly resolved, slightly decrease the truncation range and check if temperature and ITCF significantly change. If the change is significant, the data quality or spectral range may be insufficient to extract a proper normalization. The $\tau$-grid should be fine enough that any changes in it do not result in any significant change in temperature or in the ITCF. Keep the maximum of the $\tau$-grid in a range such that the unnormalized ITCF $F_u(k,0) \approx F_u(k,\tau_{max})$.
    \item[11.] If the full ITCF was sufficiently resolved, perform a linear fit on the first three data points in tau-space  $F_u(k,\tau_1 = 0), F_u(k,\tau_2), F_u(k,\tau_3)$. The linear coefficient is the slope of the ITCF $\nu$.
    \item[12.] Calculate the normalization constant A from \Cref{eq:normalisation}, where the derivative of the ITCF is given by $\nu$.
    \item[13.] Repeat the convergence check of the $\tau$-grid resolution, by increasing the resolution, repeating step 9 to 12 and monitor if A  is changing significantly depending on the resolution. 
    \item[14.] Obtain the normalized ITCF from $F(k,\tau) = A^{-1} F_{u}(k,\tau)$.
    \item[15.] Note down the total electronic static structure factor as $S_{ee}(k) = F(k, \tau)$.
\end{itemize}

\subsubsection{Rayleigh weight}

The extraction of the Rayleigh weight involves an additional step. The elastic to inelastic scattering ratio required for the Rayleigh weight can be extracted from various approaches. One possible approach is for instance to use a Chihara model \cite{Chihara_1987,Chapman2014} to obtain a fit that can reproduce the observed spectrum, that allows one to distinguish between elastic and inelastic scattering. The fit may depend on various models that may not be valid within the WDM regime and thus its parameters may be discarded. This has previously already been used for instance in Ref.~\cite{Tilo_Nature_2023} to extract the inelastic-elastic scattering ratio. With a model that reproduces the elastic scattering, one can then obtain the elastic-to-inelastic scattering ratio \cref{eq:el_inel_rat}. Since the SSF $S_{ee}(k)$ has been computed in the previous step the Rayleigh can be calculated by simply using \Cref{eq:Wr}. 

\subsubsection{Area under the ITCF}

Since the ITCF was properly normalized, a numerical integration is sufficient to obtain an accurate are below the ITCF. As a practical consideration, one can utilize the detailed-balance symmetry to make the numerical integration more accurate by rewriting the integral 
\begin{equation}
L(k) =    \int_0^{\beta} F(k,\tau) \ d\tau = 2 \int_0^{\beta/2} F(k,\tau) \ d\tau. 
\end{equation}
The comparison to the static density response function chi from another approach, for example DFT or PIMC, then yields the density. In practice, $\chi(k,0)$ should be computed from a computational approach that is sufficiently accurate within the WDM regime. 
\section{Summary and Discussion\label{sec:summary}}

We have presented a practical, largely model-free workflow to extract equilibrium and near-equilibrium plasma properties from X-ray Thomson scattering (XRTS) by working in imaginary time. The central step is to replace the ill-posed Fourier deconvolution in real frequency by a numerically stable two-sided Laplace deconvolution to obtain the imaginary-time correlation function (ITCF) $F(k,\tau)$. Within $F(k,\tau)$, detailed balance appears as a symmetry with a minimum at $\tau=\beta/2$, which enables direct and robust temperature inference. The $\tau\!\to\!0$ slope fixes the absolute normalization via the $f$-sum rule, and $F(k,0)$ provides the static electronic structure factor $S_{ee}(k)$; integrating the ITCF yields \Cref{eq:FDT}, which gives access to the static density response. Together with an elastic/inelastic intensity ratio, these quantities yield the Rayleigh weight $W_R(k)$, and comparison of $\chi(k,0)$ to single-particle or first-principles benchmarks enables density inference and model validation. In short, the imaginary-time route turns XRTS into a correlation-function metrology that emphasizes stability and interpretability. We show a basic workflow on how all these observables can be extracted in a systematic flowchart in \Cref{fig:Flowchart}.

%In practice, three ingredients are decisive for accurate results: (i) a well-characterized SIF, (ii) convergence checks with respect to the finite deconvolution window (preferably double-sided) and the $\tau$-grid resolution, and (iii) adequate signal-to-noise in the down-shifted feature to resolve detailed balance. Recent work on mosaic-crystal spectrometers and ultrahigh-resolution XRTS underscores how SIF metrology and modeling sharpen ITCF-based diagnostics \cite{Gawne2024,Gawne_PRB_2024,gawne2025heartnewxraytracing}. When these criteria are met, the ITCF pipeline is remarkably tolerant to experimental noise and yields temperatures and normalizations consistent with independent analyses \cite{Dornheim_T_2022,Thermometry2023,dornheim2023xray}.

Accurate application of the method hinges on three practical requirements: (i) a well-characterized source--instrument function (SIF), (ii) demonstrated convergence with respect to the truncated deconvolution window and the $\tau$-grid resolution, and (iii) sufficient signal-to-noise in the down-shifted feature to resolve detailed-balance asymmetry. When these conditions are met, the ITCF pipeline is tolerant to experimental noise and yields temperatures and normalizations consistent with independent analyses.

Beyond equilibrium diagnostics, the imaginary-time framework enables a straightforward test for non-equilibrium: violations of detailed balance manifest as shifts in the ITCF minima that vary with scattering vector $k$. A multi-angle strategy therefore provides a robust, first-principles indicator of relaxation dynamics in pump–probe and isochoric-heating experiments. This ability to detect departures from local thermal equilibrium (and their restoration) is a distinctive strength of the ITCF approach.

%Looking ahead, two approaches appear especially promising. First, standardizing SIF metrology and uncertainty quantification, together with automated convergence criteria for the truncated Laplace deconvolution will make ITCF-based analysis reliable across common ICF and XFEL platforms \cite{Gawne2024,Gawne_PRB_2024,gawne2025heartnewxraytracing}, while routine multi-angle collection will enable equilibrium checks and non-equilibrium studies in a single shot \cite{Bellenbaum_APL_2025}. Second, advances in analytic continuation (e.g., dual-form maximum entropy and kernel-representation methods) point toward reconstructing the full dynamic structure factor from high-quality ITCFs \cite{chuna2025dualformulationmaximumentropy,robles2025pylitreformulationimplementationanalytic}, combined with \emph{ab initio} benchmarks \cite{DornheimPRL2018,BoehmePRE2023,Bohme_PRL_2022}, opening a path to validate approximate DSFs.

Looking ahead, two directions are especially promising: (i) standardizing SIF metrology and uncertainty propagation together with automated convergence criteria for the truncated Laplace inversion, and (ii) leveraging advances in analytic continuation to reconstruct the full dynamic structure factor from high-quality ITCFs, anchored by \emph{ab initio} benchmarks.

In short, we gain stability, interpretability, and extensibility by shifting XRTS analysis to imaginary time.
Temperature, normalization,  static structure, and non-equilibrium indicators become immediately accessible so that we ultimately can uncover the complex many-body physics of WDM.
\begin{acknowledgements}
%\section*{Acknowledgments}
The work of M.P.B, D.T.B. and T.D.~was performed under the auspices of the U.S. Department of Energy by Lawrence Livermore National Laboratory under Contract No. DE-AC52-07NA27344. M.P.B., D.T.B.~and Ti.D.~were supported by Laboratory Directed Research and Development (LDRD) Grant Nos. 24-ERD-044 and 25-ERD-047.
This work was partially supported by the Center for Advanced Systems Understanding (CASUS), financed by Germany’s Federal Ministry of Education and Research (BMBF) and the Saxon state government out of the State budget approved by the Saxon State Parliament. This work has received funding from the European Research Council (ERC) under the European Union’s Horizon 2022 research and innovation programme
(Grant agreement No. 101076233, "PREXTREME"). 
Views and opinions expressed are however those of the authors only and do not necessarily reflect those of the European Union or the European Research Council Executive Agency. Neither the European Union nor the granting authority can be held responsible for them. 
This work has received funding from the European Union's Just Transition Fund (JTF) within the project \emph{R\"ontgenlaser-Optimierung der Laserfusion} (ROLF), contract number 5086999001, co-financed by the Saxon state government out of the State budget approved by the Saxon State Parliament.
W.M.M. acknowledges support from the National Science Foundation Graduate Research Fellowship Program under Grant No. DGE-2146755. This work was funded by the DOE Office of Science, Fusion Energy Science under FWP100182. We acknowledge support from Department of Energy (DOE), Office of Science, Fusion Energy Sciences, under Award No. DE-SC 0024882: IFE-STAR was issued as SLAC FWP 101126.
The work of B.A.S.~was supported by Department of Energy Award No. DE-SC0024476.
\end{acknowledgements}

%%%%%%%%%%%%%%%%%%%%%%%%%%%%%%%%%%%%%%%%%%%%%%%%%%%%%%%%%%%%%%%%%%%%%%%%%%%%%%%%
% literature
%%%%%%%%%%%%%%%%%%%%%%%%%%%%%%%%%%%%%%%%%%%%%%%%%%%%%%%%%%%%%%%%%%%%%%%%%%%%%%%%

\section*{References}
\bibliography{bibliography}

\end{document}